\documentclass{mem}
\usepackage{natbib}\usepackage{txfonts}\usepackage{balance}
\usepackage{graphicx}
\usepackage[a4paper]{hyperref}
\idline{75}{282}
\begin{document}
\def\teff{$T\rm_{eff }$}
\def\kms{$\mathrm {km~s}^{-1}$}
\def\water{H$_2$O~}

\title{
Pulsar Astrometry at the Microarcsecond Level 
}

   \subtitle{}

\author{
W.H.T.\ Vlemmings\inst{1}, S.\ Chatterjee\inst{2}, W.F.\ Brisken\inst{3}, T.J.W.\ Lazio \inst{4}, J.M.\ Cordes\inst{5}, S.E.\ Thorsett\inst{6}, W.M.\ Goss\inst{3}, E.B.\ Fomalont\inst{7}, M.\ Kramer\inst{1}, A.G.\ Lyne\inst{1}, S.\ Seagroves\inst{6}, J.M.\ Benson\inst{3}, M.M.\ McKinnon\inst{3},  D.C.\ Backer\inst{8}
\and R.\ Dewey\inst{6}
          }

  \offprints{W.Vlemmings}

\institute{
Jodrell Bank Observatory, 
The University of Manchester,
Macclesfield,
Cheshire SK11~9DL, UK
\email{wouter@jb.man.ac.uk}
\and
Jansky Fellow, National Radio Astronomy Observatory; Harvard-Smithsonian Center for Astrophysics, 60 Garden Street, Cambridge, MA 02138, USA
\and
National Radio Astronomy Observatory, P.O. Box O, Socorro, NM 87801, USA
\and
Naval Research Laboratory, 4555 Overlook Ave. SW., Washington, DC, USA
\and
Department of Astronomy and NAIC, Cornell University, Ithaca, NY 14853, USA
\and
Department of Astronomy and Astrophysics, University of California, Santa Cruz, CA 95064, USA
\and
National Radio Astronomy Observatory, 520 Edgemont Road, Charlottesville, VA 22903, USA
\and
Department of Astronomy, University of California, Berkeley, CA 94720, USA
}

\authorrunning{W.H.T.\ Vlemmings et al.}

\titlerunning{Pulsar Astrometry}

   \abstract{ Determination of pulsar parallaxes and proper motions
addresses fundamental astrophysical questions. We have recently
finished a VLBI astrometry project to determine the proper motions and
parallaxes of 27 pulsars, thereby doubling the total number of pulsar
parallaxes. Here we summarise our astrometric technique and present
the discovery of a pulsar moving in excess of 1000 \kms. As an example
of the application of high precision pulsar astrometry we also infer
the identification of 2 pulsars originating from a disrupted binary in
the Cygnus Superbubble.   }

\maketitle{}

\section{Introduction}

 Parallax and proper motion measurements obtained through
high-resolution astrometry or pulse-timing observations provide the
only model-independent distances and velocities of pulsars. Highly
accurate pulsar distances and velocities are essential for a wide
range of problems. These include:

{\bf NS Birth Sites and SNR associations:} Accurate proper
motions and parallaxes can help clarify putative pulsar--SNR
associations, leading to estimates of their true ages. Some pulsars
may be traced back to their birth sites in stellar clusters
\citep[e.g.][]{HdBdZ01} or, as discussed below, to a common origin in a disrupted binary \citep{V04}.

{\bf Galactic $n_e$:} Most pulsar distances are estimated from
their dispersion measures (DM), using a model for the Galactic
electron density \citep[e.g.][]{NE2001}. Model-independent
distances provide essential calibration points for the electron
density model, and in particular, allow much better modeling of the
local interstellar medium \citep[e.g.][]{CCL+01}.

{\bf VLBI and Interstellar Scintillation:} The
characteristics of pulsar scintillation, i.e., scintillation time and
bandwidth, depend in part upon the distribution of scattering material
along the line of sight.  Combining proper motion and parallax
measurements with interstellar scintillation observations of pulsars
allows modeling of the distribution of scattering material, and thus
contributes to modeling of the free electron density in the Galaxy.
\citet{CCL+01} demonstrate this analysis for B0919+06, from which they
find evidence for ``clumps'' with a probable scale size of 10~pc at
the edges of the Local Bubble.

{\bf NS Population Velocities:} Parallaxes and proper motions
provide model-independent transverse velocities for neutron stars
(NS), which constrain the shape of the population velocity
distribution. Pulsar velocities represent fossil information about the
evolution of close binary systems and core collapse supernovae.

{\bf Reference Frame Ties:} Astrometry on MSPs allows the
verification of solar system--extragalactic reference frame ties and
the accuracy of timing parallaxes.  The measurements also constrain
model fitting for orbital and relativistic parameters of MSPs in
binary systems.

{\bf Nuclear physics:} An accurate distance, in combination with
 observed thermal radiation from the NS surface, can be used to
 constrain the `size' of the NS photosphere, with important
 implications for the NS Equation of State.\\

Here we present a project designed to obtain accurate distances and
proper motions to a large number of pulsars to address several of the
items listed above. The results and project details are given in
Chatterjee et al.~(2005) and Brisken et al.~(2005).

\begin{figure}[]
\resizebox{\hsize}{!}{\includegraphics[clip=true]{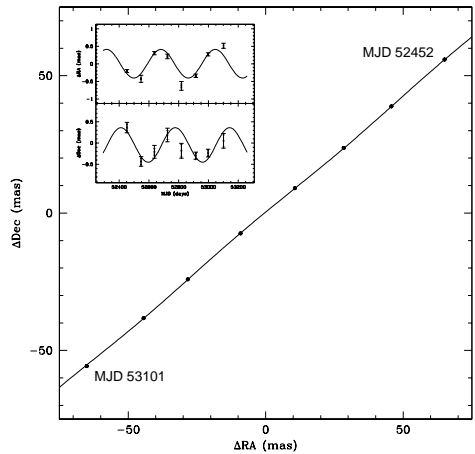}}
\caption{\footnotesize 
The motion of PSR~B1508+55 in Right Ascension and
Declination, with the best fit proper motion and parallax model
over-plotted. The error estimates for each data point are smaller
than the size of the points. The inlay shows the parallax signature of B1508+55 with the best fit proper motion subtracted.}
\label{fig1}
\end{figure}

\section{Observational Technique}

 In this project, 27 pulsars were observed with the NRAO Very Long
Baseline Array (VLBA) for 8 epochs each, spanning 2 years and over 500
hours of VLBA observing time. As a trade-off between increasing pulsar
flux at lower frequencies and improved resolution as well as reduced
ionospheric effects, all VLBA observations were conducted between 1.4
and 1.7~GHz. Observations were phase-referenced by nodding back and
forth between the target and a nodding calibrator within
$4^\circ$. For most pulsars residual calibration errors were reduced
further by employing in-beam calibration \citep{FGB+99,CCV+04},
calibration on a faint extra-galactic source within the primary VLBA
telescope beam ($\sim30'$). For the remaining pulsars we used
wide-band ionospheric calibration \citep{BBB+00}. This allows us to
reach an astrometric precision in the pulsar position often better
than 100~$\mu$arcseconds.

\section{A hyper-fast moving pulsar: B1508+55}

High pulsar velocities place stringent constraints on supernova core
collapse mechanisms. Fig.~\ref{fig1} shows the proper motion and
parallax of B1508+55, which was observed as part of our large
astrometry project. The proper motion and parallax values ($\mu_\alpha
= -73.606 \pm 0.044$~mas~yr$^{-1}$, $\mu_\delta = -62.622 \pm
0.088$~mas~yr$^{-1}$ and $\pi = 0.415 \pm 0.037$~mas) imply a pulsar
velocity with the most compact 68\% probability interval of
$1085\pm97$~\kms ($\sim1100$~\kms when corrected for differential
galactic rotation), making B1508+55 the first pulsar with such a high
measured, model independent, velocity. Constraints due to the high
velocity of B1508+55 on its formation scenario are discussed in
Chatterjee et al~(2005).

\section{Separated at birth: B2021+51 and B2020+28}

 One of the applications of high precision astrometry is the
determination of pulsar birth sites \citep[e.g.][]{HdBdZ01}. Using the
birth location information for example true kinematic ages and initial
spin periods can be determined, and pulsar birth (velocity)
distributions can be constrained.

\begin{figure}[]
\resizebox{\hsize}{!}{\includegraphics[clip=true]{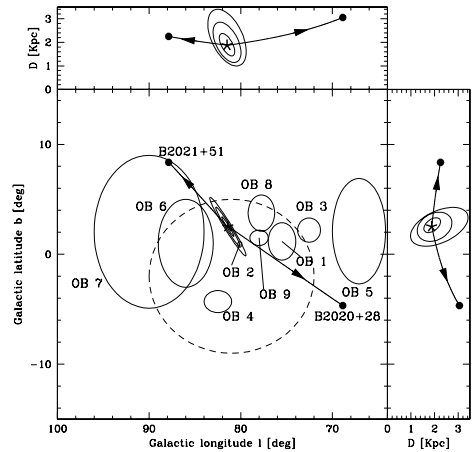}}
\caption{\footnotesize 
The 3-dimensional pulsar motion through the Galactic potential for
B2020+28 and B2021+51. The dashed circle represents the Cygnus
superbubble, while the labelled solid ellipses are the Cygnus OB
associations with positions and extents as tabulated by
\citet{UFR+01}. The extent of OB 2 is unknown and only the centre of
the association is indicated. The thick solid lines indicate the
pulsar paths, with the origin denoted by the starred symbol and the
arrows pointing in the direction of motion. The current positions are
indicated by the solid dots. The elliptical contours around the
pulsars' origin in these panels indicate the 1, 2 and 3$\sigma$
levels of the likelihood solution for the birth location. }
\label{fig2}
\end{figure}

In a sample of pulsars previously observed by the VLBA \citep{BBGT02},
we identified two pulsars that, when their orbits are calculated back
through the Galactic potential, appear to originate in a disrupted
binary in one of the Cygnus OB associations \citep{V04}. The
trajectories of these pulsars are shown in Fig.~\ref{fig2}. This
discovery implies that the kinematic age of the pulsar formed when the
second SN explosion disrupted the binary (most likely B2020+28) is
2.0~Myr (compared to the spindown age of 2.74~Myr), and the initial
spin period is $\sim200$~ms (assuming a braking index $n=3$).

Many such applications will be enabled by ongoing projects.

\begin{acknowledgements}
The National Radio Astronomy Observatory is a facility of the National
Science Foundation (NSF) operated under cooperative agreement by
Associated Universities Inc. Basic research in radio astronomy at the
NRL is supported by the Office of Naval Research.
\end{acknowledgements}

\bibliographystyle{aa}

\end{document}